\documentclass{article}
\usepackage[preprint, nonatbib]{paper_style}

\usepackage{tabularx}
\usepackage{graphicx}
\usepackage{amsmath}
\usepackage[utf8]{inputenc} 
\usepackage[T1]{fontenc}    
\usepackage[hidelinks]{hyperref}       
\usepackage{url}            
\usepackage{booktabs}       
\usepackage{amsfonts}       
\usepackage{nicefrac}       
\usepackage{microtype}      
\usepackage{xcolor}         
\usepackage{pdflscape}
\usepackage{longtable}
\usepackage{booktabs} 
\usepackage{caption}
\usepackage{multirow}

\title{Automated Treatment Planning for Interstitial HDR Brachytherapy for Locally Advanced Cervical Cancer using Deep Reinforcement Learning}

\author{
  \textbf{Mohammadamin Moradi\textsuperscript{1}} \quad
  \textbf{Runyu Jiang\textsuperscript{1,2}} \quad
  \textbf{Yingzi Liu\textsuperscript{1}} \quad
  \textbf{Malvern Madondo\textsuperscript{1}} \\
  \textbf{Tianming Wu\textsuperscript{1}} \quad
  \textbf{James J. Sohn\textsuperscript{1}} \quad
  \textbf{Xiaofeng Yang\textsuperscript{3*}} \quad
  \textbf{Yasmin Hasan\textsuperscript{1*}} \quad
  \textbf{Zhen Tian\textsuperscript{1*}}
}

\date{
\vspace{1em}
\textsuperscript{1}Department of Radiation & Cellular Oncology, University of Chicago, Chicago, IL, United States \\
\textsuperscript{2}Department of Physics, University of Chicago, Chicago, IL, United States \\
\textsuperscript{3}Department of Radiation Oncology, Emory University, Atlanta, GA, United States \\
\vspace{0.5em}
*Corresponding authors: Zhen Tian (\href{mailto:ztian@bsd.uchicago.edu}{ztian@bsd.uchicago.edu}), 
Yasmin Hasan (\href{mailto:yhasan1@bsd.uchicago.edu}{yhasan1@bsd.uchicago.edu}), 
Xiaofeng Yang (\href{mailto:Xiaofeng.yang@emory.edu}{Xiaofeng.yang@emory.edu})
}

\begin{document}
\maketitle

\begin{center}
\textsuperscript{1}Department of Radiation \& Cellular Oncology, University of Chicago, Chicago, IL, USA \\
\textsuperscript{2}Department of Physics, University of Chicago, Chicago, IL, USA \\
\textsuperscript{3}Department of Radiation Oncology, Emory University, Atlanta, GA, USA \\
\vspace{0.5em}
\textsuperscript{*}Corresponding authors: Zhen Tian (\href{mailto:ztian@bsd.uchicago.edu}{ztian@bsd.uchicago.edu}), 
Yasmin Hasan (\href{mailto:yhasan1@bsd.uchicago.edu}{yhasan1@bsd.uchicago.edu}), 
Xiaofeng Yang (\href{mailto:Xiaofeng.yang@emory.edu}{Xiaofeng.yang@emory.edu})
\end{center}


\begin{abstract}
High-dose-rate (HDR) brachytherapy plays a critical role in the treatment of locally advanced cervical cancer but remains highly dependent on manual treatment planning expertise. The objective of this study is to develop a fully automated HDR brachytherapy planning framework that integrates reinforcement learning (RL) and dose-based optimization to generate clinically acceptable treatment plans with improved consistency and efficiency. We propose a hierarchical two-stage autoplanning framework. In the first stage, a deep Q-network (DQN)-based RL agent iteratively selects treatment planning parameters (TPPs), which control the trade-offs between target coverage and organ-at-risk (OAR) sparing. The agent's state representation includes both dose-volume histogram (DVH) metrics and current TPP values, while its reward function incorporates clinical dose objectives and safety constraints, including $D_{90}$, $V_{150}$, $V_{200}$ for targets, and $D_{2cc}$ for all relevant OARs (bladder, rectum, sigmoid, small bowel, and large bowel). In the second stage, a customized Adam-based optimizer computes the corresponding dwell time distribution for the selected TPPs using a clinically informed loss function. The framework was evaluated on a cohort of patients with complex applicator geometries. The proposed framework successfully learned clinically meaningful TPP adjustments across diverse patient anatomies. For the unseen test patients, the RL-based automated planning method achieved an average score of 93.89\%, outperforming the clinical plans which averaged 91.86\%. These findings are notable given that score improvements were achieved while maintaining full target coverage and reducing CTV hot spots in most cases.
\end{abstract}

\section{Introduction}

Radiation therapy (RT) remains one of the most effective modalities for the treatment of various malignant tumors, offering high conformality and organ preservation advantages compared to surgical interventions~\cite{khan2014khan}. Among the different radiation therapy techniques, high-dose-rate (HDR) brachytherapy plays a crucial role for many cancer sites due to its ability to deliver highly localized dose distributions directly within or adjacent to the target volume while minimizing radiation exposure to surrounding healthy tissues~\cite{kubo1998high}. HDR brachytherapy is commonly employed in the management of gynecological~\cite{jiang2025deep, tian2019machine}, prostate~\cite{zhang2020automatic, mendez2018high}, head and neck~\cite{nag2001american, mazeron2002head}, and breast cancers~\cite{deng2017brachytherapy}, often as part of a multi-modality treatment strategy. In particular, for patients with locally advanced cervical cancer (LACC), HDR brachytherapy remains the standard for achieving durable local control and improving overall survival outcomes~\cite{le2022high, liu2014high}.

Brachytherapy planning is, however, a highly complex process that requires precise dwell time and source position optimization to balance target coverage and organ-at-risk (OAR) sparing. In cervical cancer brachytherapy, multiple target structures such as high-risk clinical target volume (CTV$_{\mathrm{HR}}$) and intermediate-risk clinical target volume (CTV$_{\mathrm{IR}}$), along with multiple OARs including bladder, rectum, sigmoid, small bowel, and large bowel, must be simultaneously considered. Furthermore, many patients receive external beam radiation therapy (EBRT) prior to brachytherapy, which adds an additional layer of complexity since cumulative dose constraints must account for the dose already delivered during EBRT. The dose distributions are governed by complex physical interactions that are highly patient-specific, requiring customized optimization for each individual anatomy~\cite{rivard2009evolution}. Traditional inverse planning methods, such as dose-based or graphical optimization, rely heavily on expert-defined weighting factors, iterative trial-and-error adjustments, and planner experience, which can introduce variability and inefficiencies into the clinical workflow~\cite{morton2008comparison, dinkla2015comparison}.

In recent years, artificial intelligence (AI) and machine learning (ML) techniques have emerged as powerful tools for improving various aspects of healthcare, including diagnostic imaging, treatment response prediction, and clinical decision support~\cite{vaananen2021ai, shaheen2021applications, koski2021ai}. Within radiation oncology, AI-driven algorithms have shown promise~\cite{yakar2021artificial} in several key domains such as automated contouring~\cite{court2024artificial}, image registration~\cite{teuwen2022artificial}, and quality assurance outcomes prediction~\cite{wall2020application}. Within EBRT, knowledge-based planning (KBP) models have been widely studied to predict achievable dose distributions based on prior plan libraries~\cite{ge2019knowledge, shiraishi2016knowledge}. More recently, deep learning models have shown strong performance in dose prediction for EBRT~\cite{lagedamon2024deep}. Increasingly, AI approaches are being investigated to automate radiation treatment planning (autoplanning), with the goal of generating high-quality treatment plans with reduced human intervention, improved consistency, and potential workflow acceleration. Among various machine learning paradigms, reinforcement learning (RL) offers a unique advantage for treatment planning problems due to its intrinsic formulation of sequential decision-making under uncertainty~\cite{pu2022deep}. RL models problems as Markov decision processes (MDPs), where an agent learns to optimize a policy by interacting with an environment, observing states, taking actions, and receiving rewards~\cite{sutton1998reinforcement}. This framework is well-suited to radiotherapy planning, where decisions involve iterative adjustments to planning parameters that directly impact clinical objectives, and where long-term plan quality depends on cumulative actions over multiple optimization steps. Unlike supervised learning, RL does not require explicit paired input-output datasets, making it highly adaptable to problems where optimal solutions may not be known a priori~\cite{liu2022automatic, shen2021hierarchical, shen2021improving, gao2024human}.

Despite its promise, significant challenges remain when applying AI and RL techniques to radiotherapy planning, and particularly to brachytherapy. Unlike external beam radiotherapy (EBRT), where beam geometries are relatively standardized, brachytherapy involves complex intracavitary or interstitial applicator geometries, highly patient-specific anatomy, and intricate trade-offs between multiple targets and multiple OARs. Prior work in DRL-based brachytherapy planning has often been limited to simplified scenarios, such as tandem-and-ovoid (T/O) applicator configurations with a restricted set of OARs and limited scope of dose metrics~\cite{shen2019intelligent}. Furthermore, the non-convex and discontinuous nature of brachytherapy optimization spaces presents additional difficulties for AI models to robustly learn clinically acceptable solutions across diverse patient cases.

In this study, we propose a comprehensive, fully automated framework for HDR brachytherapy treatment planning that integrates reinforcement learning with dose-based optimization. Our study advances the current literature in several important aspects. First, unlike previous studies that focused primarily on tandem-and-ovoid applicators, our dataset includes a broader and more complex cohort of locally advanced cervical cancer patients, incorporating a wide range of applicator geometries and anatomical variations. Second, we explicitly include all clinically relevant OARs (bladder, rectum, sigmoid, small bowel, and large bowel) along with having two CTVs in our optimization and learning processes, addressing a limitation of many previous works. Third, we directly model clinical dose-volume metrics used in actual clinical practice, such as $D_{90}$, $V_{100}$, $V_{150}$, and $V_{200}$ for target coverage, and $D_{2cc}$ values for all OARs, thereby closely replicating real-world clinical procedures. Fourth, we employ a customized Adam-based dwell time optimizer that minimizes a carefully designed objective function, improving both solution stability and optimization speed while allowing high-quality initialization for the RL agent. Fifth, we introduce a customized reward function that accurately reflects clinical trade-offs, balancing target coverage and OAR sparing with reward shaping techniques such as dynamic penalty and bonus mechanisms. Sixth, our reinforcement learning state representation incorporates both dose-volume histogram (DVH)-based plan quality metrics and the current treatment planning parameters (TPPs), enabling the agent to directly learn the relationship between planning parameter choices and resulting clinical outcomes. These contributions collectively enable our framework to handle the full complexity of HDR brachytherapy planning in a clinically realistic setting. To achieve this, we formulate HDR brachytherapy planning as a two-stage hierarchical decision problem. In the first stage, a deep Q-network (DQN)-based reinforcement learning agent iteratively selects optimal treatment planning parameters (TPPs), including weighting factors for targets and OAR objectives. In the second stage, a deterministic dwell time optimizer computes the corresponding dwell time distribution for the selected TPPs using a dose-based loss function. The dwell time optimizer relies on precomputed patient-specific dose influence matrices and employs automatic differentiation to efficiently compute gradients. The resulting dose distributions are evaluated via a clinically-informed reward function, providing feedback to guide RL policy learning across multiple patient cases.

The remainder of this paper is organized as follows. Section~\ref{sec:methods} describes the methods in detail, including the overall system architecture, reinforcement learning framework, dwell time optimizer design, and reward function formulation. Section~\ref{sec:results} presents the experimental results, including both training performance and plan quality evaluations across multiple patients. Section~\ref{sec:discussion} discusses the implications, limitations of the proposed framework, and future works. Finally, Section~\ref{sec:conclusion} concludes the paper.

\section{Methods} \label{sec:methods}

\begin{figure}[ht]
    \centering
    \includegraphics[width=\textwidth]{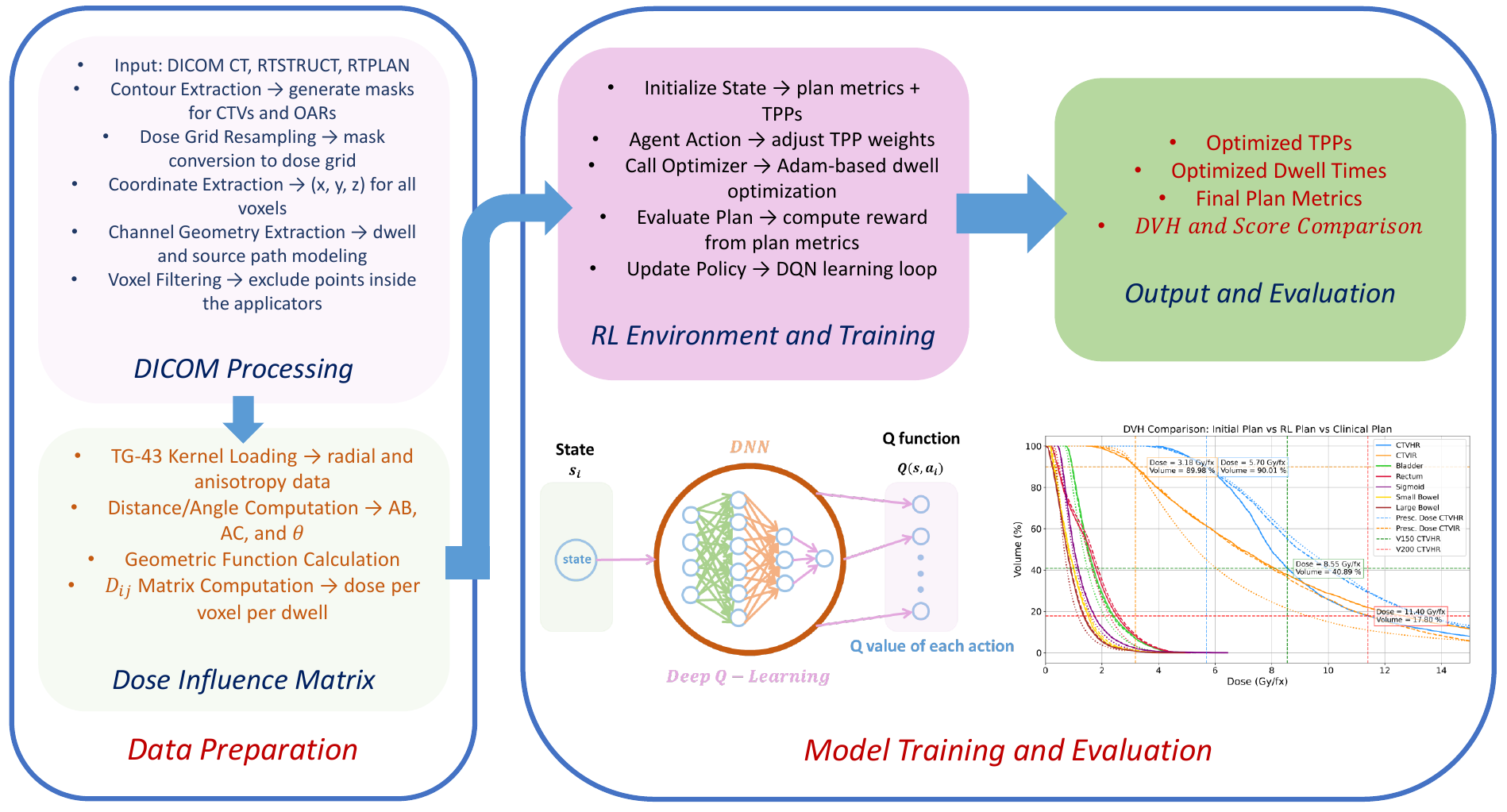}
    \caption{Workflow of the proposed RL-guided HDR brachytherapy planning system. The pipeline begins with DICOM preprocessing and anatomical mask generation, followed by coordinate extraction and exclusion of voxels near source channels. Patient-specific dose influence matrices ($D_{ij}$) are computed using TG-43-based dose modeling. The RL environment leverages plan quality metrics and treatment planning parameters (TPPs) to guide a deep Q-network (DQN) agent. At each episode, the agent adjusts TPPs, which are passed to an Adam-based dwell time optimizer. The resulting plan is evaluated via a custom reward function, and the agent’s policy is updated accordingly. The final output consists of optimized dwell times and dosimetric metrics for clinical evaluation.}
    \label{fig:method_flowchart}
\end{figure}

We developed a hybrid treatment plan optimization framework that combines DRL with dose-based optimization to generate high-quality treatment plans for HDR brachytherapy. We decomposed the HDR brachytherapy optimization problem into four sequential but interdependent stages. Figure~\ref{fig:method_flowchart} presents an overview of the proposed pipeline. The process begins with DICOM parsing and contour extraction, where anatomical structures and source channels are converted into binary masks and 3D coordinates. These are processed to exclude voxels inside and near applicators, ensuring valid dose evaluation points. In the second stage, the TG-43 formalism is used to construct patient-specific dose-influence matrices ($D_{ij}$) by calculating geometric, radial, and anisotropy components for each source–voxel pair. In the third stage, a DQN-based RL agent learns to select appropriate TPPs, which include weighting factors applied to the CTV targets and the OARs. The DQN agent interacts with a custom-designed environment that simulates the brachytherapy planning optimization. This optimization module solves for the optimal dwell time distribution and employs a dose-based loss function incorporating both target coverage metrics and OAR sparing objectives. We utilize an Adam-based gradient optimizer to efficiently converge to an optimal solution. The optimization leverages precomputed patient-specific dose influence matrices ($D_{ij}$ matrices), which directly map dwell times to voxel-wise dose distributions across multiple target volumes and OARs. Moreover, the state space, action space, and reward function are specifically designed to reflect clinically meaningful metrics and prioritize clinically acceptable dose distributions. Finally, the system outputs the optimized TPPs, dwell times, and corresponding plan quality metrics; these are then compared against clinical baselines. In the following subsections, we describe in detail the patient data processing, the deterministic dwell time optimization procedure, the design and implementation of the reinforcement learning agent, the formulation of the state and action spaces, and the reward function design. 

\subsection{Patient Data Processing} \label{sec:rl}

For each patient, the DICOM files were loaded from the raw DICOM dataset using an internal parser function. The CT resolution, voxel dimensions, and spatial origin were extracted from the CT DICOM metadata. All anatomical structures are encoded as spatial masks and aligned with physical voxel grids required for subsequent dose calculation and influence matrix generation. After generating binary structure masks on the dose grid, the next step involves identifying the real-world spatial coordinates of all voxels belonging to each anatomical structure. This step effectively converts binary spatial masks into physical coordinates needed for dose deposition modeling. 

Next, it is necessary to obtain the precise spatial coordinates of all dwell positions and the physical paths of brachytherapy applicator channels. First, the RP file was parsed to identify all active brachytherapy channels and their associated dwell positions. Each channel's physical trajectory was reconstructed by collecting its control point positions and interpolating between them to create a smooth centerline. To represent the physical volume occupied by each channel, a curved cylindrical surface was generated by sweeping a circular cross-section along the interpolated channel path. Furthermore, an accurate modeling of OAR exposure requires excluding voxels that fall within or too close to the physical path of the source channels. These points are considered invalid for dose evaluation due to overlap with the applicator hardware. 

Finally, to compute the dose-influence matrix ($D_{ij}$), we implemented the full TG-43 formalism. The final $D_{ij}$ matrices were saved as MATLAB \texttt{.mat} files and are later used as fixed input for optimization or reinforcement learning procedures.

\subsection{Reinforcement Learning} \label{sec:rl}

RL was utilized to automate the selection of TPPs for HDR brachytherapy planning. The TPP selection problem was formulated as a MDP, where at each decision point the RL agent observes the current planning state, selects an action corresponding to an adjustment in TPP values, and receives a reward reflecting the clinical quality of the resulting dose distribution. The agent learns to balance competing clinical objectives such as target coverage and OAR sparing by iteratively exploring different TPP configurations and observing the corresponding dose metrics.

We employed the DQN algorithm as implemented in the Stable-Baselines3 framework. The DQN architecture consists of a fully connected neural network with two hidden layers, each containing 128 units and ReLU activation functions. The network outputs Q-values corresponding to available actions, enabling the agent to select actions based on estimated long-term reward. Experience replay and target network updates were incorporated to stabilize training and improve sample efficiency. 

The DQN agent was trained on a cohort of seven patients, each characterized by individual anatomical geometries and dose influence matrices. During training, patient cases were randomly sampled at each episode to promote generalization across anatomical variations. Each patient’s dose matrices, fractionation scheme, prescription doses, clinical goals, and reference clinical plan metrics are preloaded based on the patient ID. These parameters define the individualized optimization problem for each case. The agent interacted with the environment for a total of 40,000 timesteps, with training progress monitored via TensorBoard logging. Upon completion of training, the final model was saved and subsequently evaluated on unseen scenarios to assess its ability to generalize to new patient anatomies.

At each interaction with the environment, the agent observes a state vector $\mathbf{s}_t \in \mathcal{S}$ that encapsulates two types of information: (i) the current dosimetric quality of the treatment plan, and (ii) the current values of the TPPs used by the dwell time optimizer. The dosimetric features include clinically relevant plan quality metrics optimized by the deterministic dwell time optimizer based on the current TPP configuration. The full state can be expressed as:

\[
\mathbf{s}_t = \left[ \mathbf{m}_t, \ \mathbf{w}_t \right]
\]
where $\mathbf{m}_t$ represents the vector of dose metrics at time step $t$, and $\mathbf{w}_t$ contains the current TPP weight values.

The action space $\mathcal{A}$ is discretized to allow the agent to incrementally adjust individual TPPs. At each decision step, the agent selects one action $a_t \in \mathcal{A}$ that modifies one specific treatment planning parameter by a fixed step size, either increasing or decreasing its value within predefined bounds. Moreover, a ``do-nothing'' action leaves all parameters unchanged, allowing the agent to terminate adjustments or maintain satisfactory solutions. In this study, a total of $17$ discrete actions were defined.

\subsection{Dwell Time Optimization} \label{sec:meth_ae}

Following the adjustment of TPPs by the RL agent, the computation of dwell times is performed by an Adam-based optimizer. Patient-specific dose influence matrices ($D_{ij}$ matrices) are precomputed for each anatomical structure and imported from external files. The optimizer minimizes a composite loss function that combines quadratic deviations from clinical goals, scaled by the TPP weights. The loss function includes:

\begin{itemize}
    \item \textbf{CTVIR coverage:}
    \[
    L_\mathrm{CTVIR} = w_{\mathrm{IR}} \cdot (D_{90}^{\mathrm{CTVIR}} - 100)^2
    \]
    
    \item \textbf{CTVHR hot spots:}
    \[
    L_{\mathrm{hot spots}} = w_{\mathrm{D50}} \cdot \left( \frac{D_{50}^{\mathrm{CTVHR}}}{150} \right)^2 
    + w_{\mathrm{D20}} \cdot \left( \frac{D_{20}^{\mathrm{CTVHR}}}{200} \right)^2
    \]

    \item \textbf{OAR sparing:}
    \[
    L_{\mathrm{OAR}} = \sum_{s} w_s \cdot \left( \frac{p_s}{100} \right)^2
    \]
    where $p_s$ is the normalized $D_{2cc}$ for each OAR relative to its clinical threshold.

    \item \textbf{Penalty terms:} Hard penalties are applied to enforce strong adherence to constraints, including deviations in $D_{90}^{\mathrm{CTVHR}}$ and excessive OAR doses exceeding thresholds:
    \[
    \mathrm{Penalty} = 100 \cdot |D_{90}^{\mathrm{CTVHR}} - 100| 
    + \sum_{s} 10 \cdot w_s \cdot \max(p_s - 100, 0)
    \]
\end{itemize}

The total loss is given by:
\[
\mathcal{L} = L_\mathrm{CTVIR} + L_\mathrm{hot spots} + L_\mathrm{OAR} + \mathrm{Penalty}.
\]

The above loss function is minimized with respect to dwell times using the Adam optimizer. To ensure positivity of dwell times, optimization is performed in the logarithmic domain. Specifically, we optimize over $\log \mathbf{d}$ (where $\mathbf{d}$ is the dwell time) and exponentiate at each iteration. At each optimization step, the clinical dose metrics from the current dwell time estimate are computed and the loss and its gradients are evaluated using automatic differentiation. Finally, log-dwell variables are updated via Adam optimization. Early stopping is employed with a patience of 500 iterations, halting optimization if no significant loss improvement ($\Delta L < 0.1$) is observed. Upon convergence, the optimizer returns the optimized dwell time vector $\mathbf{d}^*$, the final clinical dose metrics for all targets and OARs as well as the achieved loss value.

\textbf{Remark. }During this study, we systematically evaluated several gradient-based and derivative-free optimizers for this task. Conjugate Gradient (CG) was found to be unstable in our setting and frequently diverged. Limited-memory BFGS (L-BFGS) provided better stability and much faster convergence than CG, but still suffered from occasional divergence under stiff penalty configurations. CMA-ES (Covariance Matrix Adaptation Evolution Strategy), a popular evolutionary strategy, exhibited excellent stability and global convergence properties, but was prohibitively slow for our problem size and frequency of calls. In contrast, Adam consistently achieved convergence with high stability, offered competitive speed, and leveraged GPU acceleration efficiently. We speculate that Adam’s momentum-based updates and robustness to noisy gradients are well suited to the nonconvex landscape of brachytherapy optimization, which involves discontinuous penalty gradients and sparse dose contributions. The use of logarithmic dwell encoding further improves numerical conditioning, which complements Adam’s adaptive step size mechanism. These properties made Adam an ideal candidate for our proposed pipeline.

\subsection{Environment Configurations}

The RL environment is built following the OpenAI Gym interface and is fully integrated with the underlying dose-based optimizer described in the previous section. At the start of each episode, the environment randomly selects a patient from a predefined list of available patient anatomies. Initial TPPs are sampled uniformly at random within predefined parameter bounds, followed by an initial optimization to establish starting dwell times and plan metrics.

The observation space consists of a two-dimensional feature matrix combining both dosimetric and parametric information.  TPP updates are constrained within their respective bounds, and attempting to modify parameters outside their allowable range incurs a fixed penalty.

After each parameter update and re-optimization, the environment evaluates the resulting plan and computes a scalar reward. The reward is primarily based on the improvement in plan quality relative to the previous step, scaled by a factor of 100. Additionally, boundary violations in parameter updates incur fixed penalties to discourage illegal actions. Each episode proceeds for a maximum number of steps (e.g., 5 steps). An episode terminates when the agent either reaches this limit or continually selects the do-nothing action, indicating convergence.

\textbf{Remark.} This state representation, composed of plan quality metrics and treatment planning parameters, constitutes a complete and minimal encoding of the relevant clinical and optimization context for each patient. The dosimetric features summarize the current outcome of the plan in terms of target coverage, dose conformity, and OAR sparing—aligning directly with clinical evaluation criteria. Simultaneously, the TPPs encapsulate the agent’s control levers, reflecting how dwell time optimization will respond to further parameter changes. The state representation encodes the downstream behavior of the dose optimizer and the resulting plan trade-offs, abstracting away individual anatomical details while retaining all clinically actionable information. Therefore, this representation is sufficient to define the patient-specific planning landscape and necessary to capture the cause-effect relationship between planning priorities and dosimetric outcomes.

\subsection{Reward Function Design}

The reward function plays a significant role in guiding the RL agent toward generating clinically acceptable treatment plans. In this framework, the reward function is explicitly designed to evaluate plan quality based on multiple competing clinical objectives. This function transforms dose metrics into a scalar reward via a multi-stage scoring process that integrates both goal proximity and clinical safety margins.

The reward function operates by assigning scores to each dosimetric metric based on its deviation from clinical goals. For each metric, the following elements are defined:

\begin{itemize}
    \item A target goal value.
    \item A tolerance window that defines acceptable deviations.
    \item A weighting factor that controls the relative contribution of each metric to the total score.
    \item A directionality parameter indicating whether higher, lower, or target-matching values are preferred.
\end{itemize}

The scoring logic is implemented through a generalized scoring function $S(\cdot)$:
\begin{itemize}
    \item For \textbf{higher-is-better} metrics (direction: ``higher''):
    \[
    S = 
    \begin{cases}
    w & \text{if } v \geq \text{goal},\\
    w \cdot \frac{v - (\text{goal} - \text{tol})}{\text{tol}} & \text{if } v \in [\text{goal} - \text{tol}, \text{goal}],\\
    0 & \text{otherwise.}
    \end{cases}
    \]
    
    \item For \textbf{lower-is-better} metrics (direction: ``lower''):
    \[
    S = 
    \begin{cases}
    w & \text{if } v \leq \text{goal},\\
    w \cdot \frac{(\text{goal} + \text{tol}) - v}{\text{tol}} & \text{if } v \in [\text{goal}, \text{goal} + \text{tol}],\\
    0 & \text{otherwise.}
    \end{cases}
    \]

    \item For \textbf{target-centered} metrics (direction: ``target''):
    \[
    S = 
    \begin{cases}
    w \cdot \left(1 - \frac{|v - \text{goal}|}{\text{tol}}\right) & \text{if } |v - \text{goal}| \leq \text{tol},\\
    0 & \text{otherwise.}
    \end{cases}
    \]
\end{itemize}

To further incentivize aggressive OAR sparing beyond threshold values, an additional bonus is applied to OAR scores. This bonus encourages the RL agent to seek OAR doses well below clinical limits whenever feasible. The final scalar reward is computed as:
\[
R = \frac{\sum_{i} (S_i + \text{Bonus}_i)}{\sum_{i} (w_i + \text{Bonus}_\mathrm{max})} \times 100.
\]
This normalization yields a reward value scaled between $0$ and $100$ and is needed for consistent interpretation across patients and episodes.

\textbf{Remark.} All scoring terms directly reflect standard clinical dose constraints, ensuring that the learned policy prioritizes clinically meaningful trade-offs. Moreover, the tolerance windows ensure that small deviations from goals result in proportional changes in reward, which helps stabilize RL training. Utilizing reward shaping, the bonus terms encourage solutions that achieve lower-than-required OAR doses, promoting safer plans. Finally, since clinical thresholds for OARs are patient-specific, the reward function dynamically adapts to each patient anatomy.

\section{Results}  \label{sec:results}

The proposed DRL-based HDR brachytherapy autoplanning framework was implemented in Python using the Stable-Baselines3 library and PyTorch for optimizer design. All experiments were performed on a workstation equipped with an NVIDIA RTX A5000 GPU, enabling efficient parallelization of both RL training and dose optimization procedures. The deep Q-network (DQN) agent was trained for a total of 40,000 timesteps using a replay buffer size of 30,000, batch size of 512, learning rate of $10^{-4}$, and an $\epsilon$-greedy exploration strategy.

Table~\ref{tab:train_patients} and Table~\ref{tab:test_patients} summarizes patient-specific information and plan goals used during optimization for all evaluated cases. For each patient, the prescription doses for CTV$_{\mathrm{HR}}$ and CTV$_{\mathrm{IR}}$ were determined based on institutional protocols and delivered over 5 fractions. Importantly, the organ-at-risk (OAR) dose constraints for each patient were individualized based on their previously delivered EBRT doses, taking into account the cumulative dose already received by each OAR prior to brachytherapy. This individualized constraint setting ensures that the brachytherapy optimization fully respects total biological dose limits accumulated across the full course of treatment. In addition, clinically achieved OAR doses from the reference clinical plans were also included for each patient to allow direct comparison with the plans generated by the proposed framework. 

The dataset consists of 16 locally advanced cervical cancer patients in total, with the first 7 patients used for training and the remaining 9 patients reserved for testing and evaluation. This cohort includes a wide range of anatomical and dosimetric characteristics, which is essential for ensuring that the proposed optimization framework generalizes well across clinically relevant scenarios. The prescription doses for CTV$_{\mathrm{HR}}$ span from 3.23 Gy/fx to 6.20 Gy/fx, and for CTV$_{\mathrm{IR}}$ from 2.13 Gy/fx to 4.99 Gy/fx, reflecting variation in tumor size, extent, and prior EBRT contributions. Similarly, the target-specific $D_{2cc}$ thresholds for OARs such as the bladder, rectum, sigmoid, small bowel, and large bowel vary substantially between patients, capturing individualized anatomical constraints and clinical tolerances.

\begin{table}[ht]
\centering
\caption{Patient specific information for training set (Patients 1–7).}
\label{tab:train_patients}
\scriptsize
\begin{tabular}{l|ccccccc}
\toprule
\textbf{Structure} & \textbf{P1} & \textbf{P2} & \textbf{P3} & \textbf{P4} & \textbf{P5} & \textbf{P6} & \textbf{P7} \\
\midrule
CTV$_{\mathrm{HR}}$ Presc. Dose (Gy/fx)  & 6.00 & 6.20 & 5.50 & 5.79 & 3.23 & 5.55 & 5.58 \\
CTV$_{\mathrm{IR}}$ Presc. Dose (Gy/fx)  & 3.11 & 4.40 & 3.97 & 3.41 & 2.13 & 4.03 & 3.12 \\
Target Bladder $D_{2cc}$ (Gy/fx)    & 5.40 & 4.96 & 5.04 & 5.28 & 4.74 & 5.23 & 5.38 \\
Target Rectum $D_{2cc}$ (Gy/fx)     & 4.08 & 3.65 & 3.99 & 2.82 & 3.45 & 3.88 & 4.20 \\
Target Sigmoid $D_{2cc}$ (Gy/fx)    & 3.07 & 3.67 & 3.12 & 2.75 & 1.79 & 4.06 & 4.08 \\
Target Small Bowel $D_{2cc}$ (Gy/fx)& 3.61 & 2.60 & 1.83 & 2.96 & 2.14 & 2.86 & 2.78 \\
Target Large Bowel $D_{2cc}$ (Gy/fx)& 4.08 & 3.16 & 2.55 & 2.48 & 2.48 & 3.51 & 3.26 \\
Clinical Bladder $D_{2cc}$ (Gy/fx)  & 3.70 & 3.89 & 3.94 & 3.05 & 2.72 & 3.58 & 4.53 \\
Clinical Rectum $D_{2cc}$ (Gy/fx)   & 2.23 & 3.65 & 2.82 & 2.35 & 1.54 & 2.96 & 3.80 \\
Clinical Sigmoid $D_{2cc}$ (Gy/fx)  & 3.01 & 3.11 & 2.78 & 2.13 & 1.00 & 2.83 & 3.29 \\
Clinical S. Bowel $D_{2cc}$ (Gy/fx) & 1.93 & 1.68 & 1.13 & 0.71 & 1.07 & 1.03 & 3.52 \\
Clinical L. Bowel $D_{2cc}$ (Gy/fx) & 1.86 & 2.06 & 0.85 & 0.74 & 0.73 & 0.73 & 1.92 \\
Clinical CTV$_{\mathrm{HR}}$ $V_{150}$ (\%)  & 55.20 & 32.90 & 41.80 & 35.80 & 39.80 & 35.40 & 45.00 \\
Clinical CTV$_{\mathrm{HR}}$ $V_{200}$ (\%)  & 30.60 & 15.20 & 18.90 & 11.10 & 13.00 & 14.20 & 18.60 \\
\bottomrule
\end{tabular}
\end{table}

\begin{table}[ht]
\centering
\caption{Patient specific information for testing set (Patients 8–16).}
\label{tab:test_patients}
\scriptsize
\begin{tabular}{l|ccccccccc}
\toprule
\textbf{Structure} & \textbf{P8} & \textbf{P9} & \textbf{P10} & \textbf{P11} & \textbf{P12} & \textbf{P13} & \textbf{P14} & \textbf{P15} & \textbf{P16} \\
\midrule
CTV$_{\mathrm{HR}}$ Presc. Dose (Gy/fx)  & 4.86 & 5.51 & 5.86 & 4.94 & 5.70 & 5.35 & 5.18 & 4.85 & 4.65 \\
CTV$_{\mathrm{IR}}$ Presc. Dose (Gy/fx)  & 3.28 & 3.10 & 3.45 & 3.05 & 3.18 & 4.99 & 4.59 & 3.29 & 3.91 \\
Target Bladder $D_{2cc}$ (Gy/fx)    & 5.50 & 5.26 & 5.29 & 5.29 & 5.29 & 5.33 & 5.25 & 5.35 & 5.32 \\
Target Rectum $D_{2cc}$ (Gy/fx)     & 4.34 & 4.04 & 4.10 & 4.05 & 4.09 & 4.16 & 4.11 & 4.16 & 3.51 \\
Target Sigmoid $D_{2cc}$ (Gy/fx)    & 4.34 & 4.02 & 4.09 & 4.20 & 4.06 & 3.75 & 2.96 & 4.16 & 3.51 \\
Target Small Bowel $D_{2cc}$ (Gy/fx)& 3.40 & 3.14 & 3.07 & 3.19 & 2.84 & 3.14 & 2.01 & 3.19 & 3.06 \\
Target Large Bowel $D_{2cc}$ (Gy/fx)& 3.89 & 3.54 & 3.62 & 3.74 & 3.59 & 3.25 & 2.66 & 3.70 & 3.60 \\
Clinical Bladder $D_{2cc}$ (Gy/fx)  & 3.11 & 4.35 & 4.13 & 2.56 & 3.56 & 3.58 & 4.17 & 3.08 & 4.12 \\
Clinical Rectum $D_{2cc}$ (Gy/fx)   & 2.32 & 3.78 & 1.64 & 2.14 & 3.00 & 3.30 & 3.74 & 2.06 & 2.71 \\
Clinical Sigmoid $D_{2cc}$ (Gy/fx)  & 1.30 & 1.09 & 1.98 & 1.31 & 2.94 & 2.08 & 0.69 & 0.33 & 1.34 \\
Clinical S. Bowel $D_{2cc}$ (Gy/fx) & 1.78 & 1.63 & 1.01 & 0.88 & 2.68 & 3.55 & 0.83 & 0.23 & 2.59 \\
Clinical L. Bowel $D_{2cc}$ (Gy/fx) & 1.72 & 2.15 & 1.08 & 1.56 & 2.62 & 1.13 & 1.62 & 0.16 & 2.67 \\
Clinical CTV$_{\mathrm{HR}}$ $V_{150}$ (\%)  & 34.60 & 43.30 & 53.80 & 35.70 & 57.90 & 60.91 & 57.43 & 32.67 & 60.87 \\
Clinical CTV$_{\mathrm{HR}}$ $V_{200}$ (\%)  & 10.50 & 15.50 & 33.90 & 15.10 & 29.80 & 32.69 & 27.60 & 12.62 & 37.78 \\
\bottomrule
\end{tabular}
\end{table}

The training set (Patients 1–7) includes both high-dose and low-dose cases (e.g., Patient 2 with a 6.20 Gy/fx CTV$_{\mathrm{HR}}$ prescription versus Patient 5 with only 3.23 Gy/fx), as well as a range of OAR constraint tightness. For instance, Patient 4 exhibits relatively strict rectal and sigmoid constraints, whereas Patient 1 allows more generous OAR doses. Clinical dose distributions also show considerable heterogeneity: CTV$_{\mathrm{HR}}$ $V_{150}$ ranges from 32.9\% to 55.2\% and $V_{200}$ from 11.1\% to 30.6\%, indicating different levels of target dose conformity and hot spots achieved in practice. This diversity allows the RL agent to learn policies that balance trade-offs under a variety of constraint structures and reward gradients.

The testing set (Patients 8–16) introduces further variability beyond what is seen in training. These include cases with lower-than-average prescription doses (e.g., Patient 16 with CTV$_{\mathrm{HR}}$ at 4.65 Gy/fx), patients with high clinical OAR doses (e.g., Patient 9 with rectal D$_{2cc}$ of 3.78 Gy/fx), and patients exhibiting extreme high-dose volumes (e.g., Patient 16 with $V_{200}$ of 37.78\%). The presence of such diverse test cases ensures that performance improvements are not merely memorization of training patterns but reflect the ability to adapt to new anatomical layouts and constraint profiles. Therefore, achieving high-quality plans for this test set provides strong evidence that the proposed method generalizes effectively across a broad range of clinical scenarios.

Table~\ref{tab:plan_comparison_test} presents a quantitative comparison of clinical metrics after DRL optimization for all evaluated patients as well as the clinically achieved plans. Overall, the results demonstrate that the method consistently maintains target coverage while achieving substantial improvements in dose conformity and OAR sparing in both seen and unseen patients.

For CTV$_\mathrm{HR}$, the $D_{90}$ coverage target of 100\% was exactly achieved for all patients, both in clinical and DRL-optimized plans, confirming that the proposed system does not compromise on primary target coverage. More critically, the DRL optimization significantly reduced the high-dose volumes within CTV$_\mathrm{HR}$. Across the testing set, the $V_{150}$ was reduced from 49.04\% to 44.05\%, and $V_{200}$ from 23.85\% to 17.68\%, with even larger improvements in certain cases (e.g., Patient 12: $V_{200}$ reduced from 29.80\% to 19.28\%). These reductions suggest better dose conformity, potentially lowering toxicity risk while maintaining effective tumor control.

For CTV$_\mathrm{IR}$, the $D_{90}$ metric was preserved within 0.2\% of the 100\% target across all patients, confirming the stability of the optimizer in balancing multi-target coverage.

Across the cohort, improvements in overall reward scores were modest but consistent. For testing patients, the average score improved from 91.86\% (clinical) to 93.89\% (DRL). These findings are notable given that score improvements were achieved while maintaining full target coverage and reducing CTV hot spots in most cases. Importantly, the model generalized well to the test set, maintaining performance even for cases with extreme anatomical or dosimetric characteristics (e.g., Patient 15 with very low sigmoid and large bowel doses, or Patient 13 with elevated bowel doses).

\begin{table}[ht]
\centering
\caption{Plan quality metrics after RL-guided optimization compared with clinical plans for testing set patients (P8–P16). The percentage indicating each OAR's $D_{2CC}$ is relative to its clinical threshold.}
\label{tab:plan_comparison_test}
\scriptsize
\begin{tabular}{l|ccccccccc}
\toprule
\textbf{Metric} & \textbf{P8} & \textbf{P9} & \textbf{P10} & \textbf{P11} & \textbf{P12} & \textbf{P13} & \textbf{P14} & \textbf{P15} & \textbf{P16} \\
\midrule

Clinical CTV$_{\mathrm{HR}}$ $D_{90}$  (\%) & 100.00 & 100.00 & 100.00 & 100.00 & 100.00 & 100.00 & 100.00 & 100.00 & 100.00 \\
DRL CTV$_{\mathrm{HR}}$ $D_{90}$  (\%) & 100.00 & 100.00 & 100.00 & 100.00 & 100.00 & 100.00 & 100.00 & 100.00 & 100.00 \\
\hline

Clinical CTV$_{\mathrm{HR}}$ $V_{150}$ (\%) & 34.60 & 45.75 & 55.75 & 35.70 & 57.90 & 60.91 & 57.43 & 32.67 & 60.87 \\
DRL CTV$_{\mathrm{HR}}$ $V_{150}$ (\%) & 34.41 & 36.69 & 54.70 & 34.20 & 43.82 & 53.08 & 45.17 & 38.61 & 55.83 \\
\hline

Clinical CTV$_{\mathrm{HR}}$ $V_{200}$ (\%) & 10.50 & 14.68 & 33.90 & 15.10 & 29.80 & 32.69 & 27.60 & 12.62 & 37.78 \\
DRL CTV$_{\mathrm{HR}}$ $V_{200}$ (\%) & 3.05 & 8.83 & 36.19 & 8.23 & 19.28 & 27.84 & 17.76 & 3.71 & 34.23 \\
\hline

Clinical CTV$_{\mathrm{IR}}$ $D_{90}$  (\%) & 100.00 & 100.00 & 100.00 & 100.00 & 100.00 & 100.00 & 100.00 & 100.00 & 100.00 \\
DRL CTV$_{\mathrm{IR}}$ $D_{90}$  (\%) & 100.01 & 100.00 & 100.00 & 99.92 & 99.84 & 99.95 & 100.08 & 100.00 & 100.06 \\
\hline

Clinical Bladder $D_{2cc}$ (\%) & 56.54 & 82.69 & 78.07 & 48.39 & 67.29 & 67.09 & 79.51 & 57.61 & 77.40 \\
DRL Bladder $D_{2cc}$ (\%) & 70.53 & 73.40 & 89.72 & 68.15 & 73.19 & 60.40 & 68.84 & 51.13 & 77.96 \\
\hline

Clinical Rectum $D_{2cc}$ (\%) & 53.45 & 93.56 & 40.00 & 52.83 & 73.34 & 79.44 & 90.97 & 49.59 & 77.12 \\
DRL Rectum $D_{2cc}$ (\%) & 80.12 & 95.06 & 63.60 & 75.83 & 96.25 & 90.98 & 98.49 & 73.42 & 81.62 \\
\hline

Clinical Sigmoid $D_{2cc}$ (\%) & 29.95 & 27.11 & 48.41 & 31.19 & 72.41 & 55.57 & 23.30 & 8.14 & 38.15 \\
DRL Sigmoid $D_{2cc}$ (\%) & 30.20 & 29.19 & 58.37 & 50.18 & 79.56 & 59.58 & 26.97 & 4.86 & 44.50 \\
\hline

Clinical S. Bowel $D_{2cc}$ (\%) & 52.35 & 51.91 & 32.89 & 27.58 & 94.36 & 113.17 & 41.27 & 7.30 & 84.52 \\
DRL S. Bowel $D_{2cc}$ (\%) & 55.37 & 50.45 & 33.40 & 46.89 & 94.60 & 119.82 & 43.58 & 1.88 & 99.96 \\
\hline

Clinical L. Bowel $D_{2cc}$ (\%) & 44.21 & 60.73 & 29.83 & 41.71 & 72.98 & 34.75 & 61.03 & 4.46 & 74.12 \\
DRL L. Bowel $D_{2cc}$ (\%) & 40.86 & 63.29 & 22.98 & 46.26 & 72.65 & 37.73 & 70.64 & 0.67 & 86.35 \\
\hline

Clinical Score (\%) & 100.00 & 97.37 & 91.07 & 100.00 & 86.87 & 78.31 & 87.06 & 100.00 & 86.09 \\
DRL Score (\%) & 99.44 & 97.98 & 88.64 & 99.78 & 95.17 & 81.37 & 97.46 & 99.97 & 85.23 \\
\bottomrule
\end{tabular}
\end{table}

\begin{figure}[ht]
    \centering
    \includegraphics[width=\textwidth]{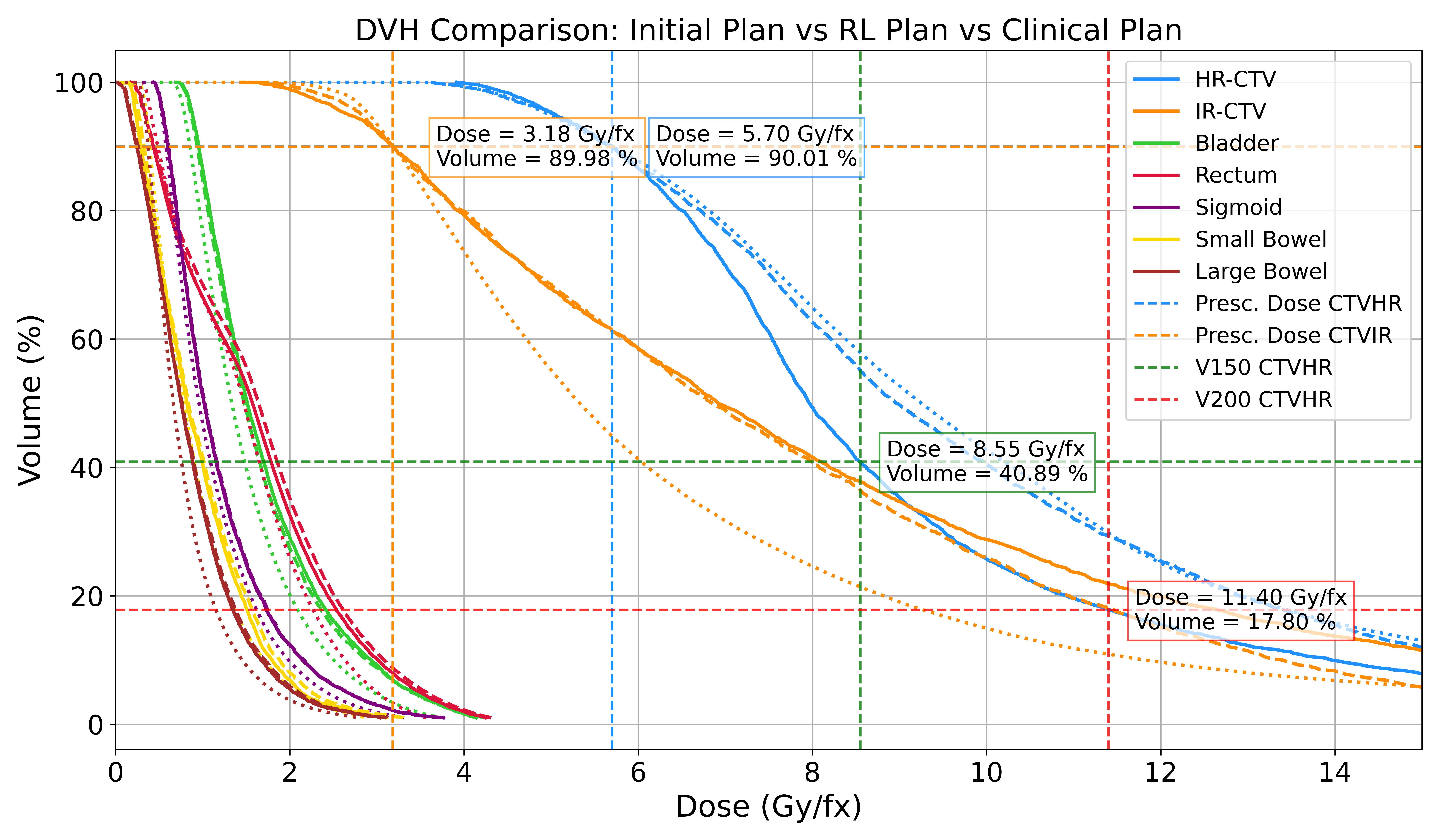}
    \caption{Dose–volume histogram (DVH) comparison for Patient 12, showing clinical plan (dotted), initial plan (dashed), and DRL-optimized plan (solid). Prescription dose lines are shown for CTV$_\mathrm{HR}$ and CTV$_\mathrm{IR}$, along with guideline thresholds for V$_{150}$ and V$_{200}$. Annotations indicate key dose and volume metrics.}
    \label{fig:dvh_p12}
\end{figure}

Figure~\ref{fig:dvh_p12} illustrates the dose–volume histogram (DVH) comparison for Patient 12, revealing key differences between the clinical, initial, and DRL-optimized plans. CTV$_\mathrm{HR}$ coverage remains preserved across all plans, with the DRL solution achieving a $D_{90}$ of 100\%, identical to the clinical target. However, the DRL plan significantly reduces excessive dose exposure, lowering the $V_{200}$ from 29.80\% to 19.28\%, and $V_{150}$ from 57.90\% to 43.82\%. These improvements are clearly visible as the DRL curve (solid blue) descends more steeply past the high-dose threshold lines, indicating improved conformity. OARs such as bladder, rectum, and bowel also maintain acceptable dose distributions, with minor trade-offs. The overall plan quality score for Patient 12 improved from 86.87\% in the clinical plan to 95.17\% in the DRL plan. The DVH thus visually confirms that the DRL-generated solution achieves superior dose shaping and high plan quality.

\section{Discussion} \label{sec:discussion}

Across 16 patients, the proposed method consistently maintained target coverage, improved dose conformity, and offered clinically competitive trade-offs for OAR sparing. A key result of this work is the significant reduction in CTV$_\mathrm{HR}$ high-dose volumes (V$_{150}$ and V$_{200}$), achieved without compromising D$_{90}$ coverage. This suggests that the RL agent has learned to shape dose distributions more conformally by adjusting TPP weights. Performance on test patients, many of whom presented with anatomical and dosimetric conditions not seen during training, confirms the model’s generalization capability. For instance, the DRL agent managed to produce high-quality plans for patients with extreme OAR constraints, low prescription doses, or highly variable high-dose volumes. The diversity and success of these test cases suggest that this framework may be extensible to a broader patient population, particularly when combined with adaptive learning or patient-specific tuning.

Nevertheless, several areas remain open for future research. Incorporating 3D convolutional encodings or transformer-based attention mechanisms may allow the RL agent to perceive spatial patterns that are currently abstracted away by scalar inputs. Additionally, the reward function, although clinically informed and differentiable, could benefit from task-specific fine-tuning or even adversarial training schemes to penalize implausible dose shapes. An exciting avenue is the integration of large language model (LLM)-guided reward shaping or expert-informed reward functions derived from clinical guidelines and textual reports, which could encode implicit decision-making rules into the learning process. In parallel, inverse reinforcement learning (IRL) could be used to infer optimal behavior directly from expert plans, thereby grounding the reward function in real-world clinical practice. Another extension is to move beyond discrete action spaces: a continuous action formulation could enable more fine-grained parameter updates and smoother exploration, potentially improving learning stability and efficiency. Finally, expanding the agent’s scope beyond TPP tuning—such as learning dwell position activation patterns or even suggesting applicator or needle placements—could transform the system into a comprehensive, end-to-end planning solution. 

Finally, to support clinical translation, future work must emphasize interpretability and physician-in-the-loop interfaces. Reinforcement learning agents must not only perform well but also be able to explain their rationale and allow manual override when needed. Integration with commercial TPS systems, validation with more datasets, and physician scoring evaluation will be essential to elevate this line of research from proof-of-concept to clinical readiness.

\section{Conclusion} \label{sec:conclusion}

In this study, we developed and evaluated a fully automated, reinforcement learning-driven treatment planning framework for HDR brachytherapy in locally advanced cervical cancer. The proposed system integrates a DQN-based agent with a deterministic, Adam-based dose optimizer to automatically generate clinically acceptable treatment plans while replicating real-world clinical decision-making processes. Unlike prior studies that have focused on simplified applicator geometries or limited subsets of clinical objectives, our framework incorporates comprehensive anatomical modeling, including all relevant OARs, and directly integrates clinical dose-volume metrics such as $D_{90}$, $V_{150}$, $V_{200}$, and $D_{2cc}$ into both the optimization objective and the reinforcement learning reward function. The agent’s state representation, which combines both plan quality metrics and TPPs, allows for more effective learning of complex dosimetric trade-offs. The results demonstrate that the proposed system can successfully learn optimal parameter adjustment strategies across a diverse cohort of patients, producing high-quality treatment plans that achieve appropriate target coverage and maintain OAR doses within clinical tolerance levels. The use of a customized reward function and optimizer contributed to both learning stability and plan quality consistency, addressing key challenges inherent to HDR brachytherapy planning.

\begin{ack}

This project is partially supported by the National Institute of Health under Award Number R37CA272755 and R01CA272991.

\end{ack}

\subsection*{Code Availability}

The code is provided upon request.

\bibliography{ref.bib}
\end{document}